\documentclass[preprint,12pt]{elsarticle}
\usepackage{graphicx}
\usepackage{amssymb}
\usepackage{amsmath}

\journal{Nuclear Physics A}

\begin{document}

\begin{frontmatter}

\title{Antineutron and antiproton nuclear interactions at very low energies}
\author[a]{E.~Friedman}
\cortext[cor1]{elifried@cc.huji.ac.il}
\address[a]{Racah Institute of Physics, The Hebrew University, 91904
Jerusalem, Israel}

\begin{abstract}
Experimental annihilation cross sections of antineutrons and antiprotons
at very low energies are compared. Features of Coulomb focusing are
observed for $\bar p$ annihilation on protons. 
Direct comparisons for heavier targets
are not straightforward due to lack of overlap between 
targets and energies of experimental
results for $\bar p$ and $\bar n$. Nevertheless, the
annihilation cross sections for $\bar n$ on nuclei cannot be described 
by an optical potential that fits well all the available data on $\bar p$
interactions with nuclei. Comparisons made with the help of this potential
reveal in the $\bar n$ data features similar to Coulomb focusing.
Direct comparisons between $\bar n$ and $\bar p$ annihilations at very
low energies would be possible when $\bar p$ cross sections are measured
on the same targets and at the same energies as the available cross sections
for $\bar n$. Such measurements may be possible in the foreseeable future.

\end{abstract}

\begin{keyword}
antineutron-nucleus and antiproton nucleus interactions.
\end{keyword}

\end{frontmatter}

\section{Introduction}
\label{sec:intro}

Experimental results for annihilation cross sections of antineutrons on
several nuclei at very low momenta across the periodic table have been
published by the OBELIX collaboration more than a decade ago \cite{ABB02}.
Exceedingly large cross sections have been reported below 180 MeV/c
and features typical of reactions taking place at the nuclear
surface have been noted. However, quantitative analyses of those data in
terms of optical potentials have not been reported. 
In contrast, data for antiproton-nucleus 
interaction both below threshold (antiprotonic atoms) and above have been
repeatedly analysed and led to fully consistent quantitative picture of the
interaction at low energies \cite {FGa07}.

Comparing between $\bar p$ and $\bar n$ interactions with nuclei, it is
unfortunate that experimental results are available mostly for
different nuclei at different energies so that direct comparisons
are not possible. Only for $\bar p$ and $\bar n$ interaction with
the proton is it possible to compare directly experimental results.
For nuclei we adopt an optical model as a tool for
comparisons between different results. Optical potentials simply
related to nuclear densities have a long history 
in nuclear physics of smoothly
interpolating values of observables over energy and 
over atomic and mass numbers
\cite{GPT68}.  It is shown in the present work that such an approach
which is successful for $\bar p$-nucleus
interactions poses problems when applied to antineutrons. 

In sec.\ref{sec:CoulF} we compare total annihilation cross sections 
for $\bar p p$  and $\bar n p$ at very low energies and describe
the mechanism of Coulomb focusing which is responsible for
large differences at low  energies.
In sec.\ref{sec:pbarsNUC} we re-examine results for 
antiproton-nucleus interactions
at low energies showing the high degree of consistency between the
various experimental results along the periodic table, both 
below and above threshold. 
In sec.\ref{sec:nbarsNUC} we confront the experimental annihilation 
cross sections for $\bar n$ on nuclei with various calculations. 
The increased importance at low energies of Coulomb focusing 
is discussed in some detail.

 Section \ref{sec:summ} is a discussion
where it is proposed to match the existing data 
of annihilation cross sections
for $\bar n$ on nuclei by measuring annihilation cross sections for
$\bar p$ on the same nuclei at the corresponding energies in an attempt
to shed light on what appears to be a puzzle. Such measurements should
be feasible in the foreseeable future.

\section{Antiproton and antineutron annihilation on the proton}
\label{sec:CoulF}

\begin{figure}[t]
 \begin{center}
  \includegraphics[height=8cm,width=0.8\linewidth]{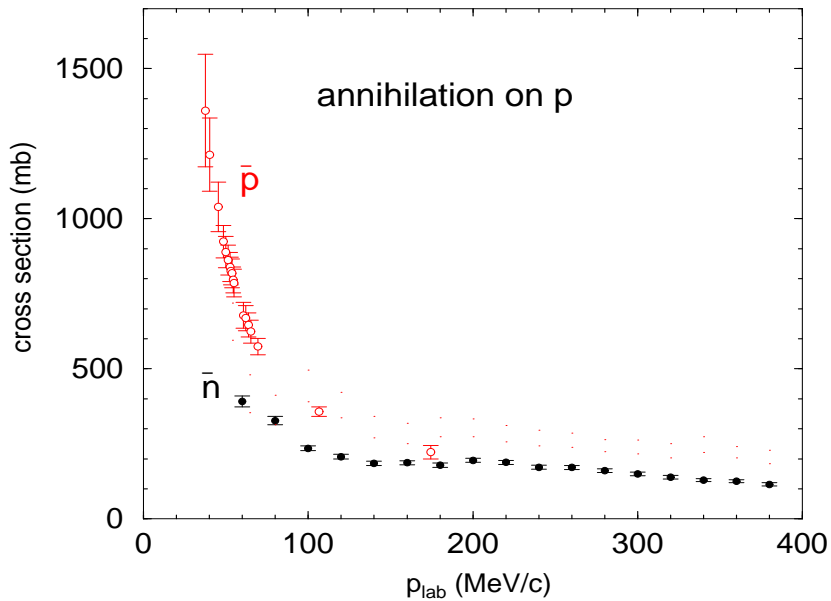}
\caption{Total annihilation cross sections 
on the proton, from \cite{BBC96,ZBB99,Ben97,BBC97}. Open circles for 
$\bar p$, filled circles for $\bar n$.}
  \label{fig:pbarnbarp}
 \end{center}
\end{figure}

The OBELIX collaboration measured
total annihilation cross sections for antiprotons on the proton 
 \cite{BBC96,ZBB99,Ben97} and for antineutrons on the proton
\cite{BBC97}. From the results shown in fig.\ref{fig:pbarnbarp}
it is seen that whereas the $\bar p$ and $\bar n$ 
cross sections tend to be very close 
to each other above 200 MeV/c very large differences appear below
$\approx$ 80 MeV/c. The $\bar n$ cross sections show moderate increase
of cross sections as the energy goes down, most likely due to the
expected $1/v$ dependence of the $s$-wave cross section. However the increase
of the cross sections for the $\bar p$ is much stronger
than the increase for $\bar n$. This is
the result of the so-called Coulomb focusing effect which has already 
been observed in annihilation cross sections of $\bar p$ on nuclei
\cite{BFG01}. In situation of very strong absorption, typical of
antiproton interactions, the `black disk' cross section $\pi R^2$
is replaced by
\begin{equation}
\label{eq:CF}
\sigma_R=\pi R^2 (1+\frac{2MZe^2}{\hbar^2k_{\rm L} kR})
\end{equation}
with M the mass of the proton, $k$ and $k_{\rm L}$ the cm and lab wave numbers,
respectively and $R$ the black disk radius. For very low energies 
(and momenta) the second term in the brackets becomes dominant. Note 
that the $Z/R$ dependence increases very rapidly along the periodic
table, considering that R changes with A$^{1/3}$, with A the mass 
number of the nucleus. It is therefore expected that annihilation
cross sections for $\bar p$ on nuclei will increase as energy is lowered 
much faster than do the corresponding cross sections for antineutrons.

\section{Antiproton-nucleus interaction}
\label{sec:pbarsNUC}

With energies close to threshold, we begin with antiprotonic
atoms, where the experimental results of the PS209
collaboration at CERN \cite{TJC01} provided 
 high-quality data for several sequences of isotopes along the periodic
table.  Detailed analyses of these results  have been
published in a series of papers, dedicated each to a particular
subset of the data such as neighboring nuclei or isotopes of the same
element. In the present context  we discuss only  {\it global} fits
of optical potentials
to the entire set of 90 data points relating to strong interaction level
shifts and widths in $\bar p$ atoms \cite{FGM05}.

In line with other types of exotic atoms \cite{FGa07}, 
the interaction of
antiprotons with nuclei at threshold is described in terms of an optical
potential, which in the simplest $t \rho$ form is given by
\begin{equation}
\label{eq:pbarspotl}
2\mu V_{{\rm opt}}(r) = -4\pi(1+\frac{\mu}{M}
\frac{A-1}{A})[b_0(\rho_n+\rho_p)
  +b_1(\rho_n-\rho_p)]~~,
\end{equation}
where $\mu$ is the reduced mass of the $\bar p$,
 $\rho_n$ and $\rho_p$ are the neutron and proton density
distributions normalized to the number of neutrons $N$ and number
of protons $Z$, respectively, $A=N+Z$, and $M$ is the mass of the
nucleon. The complex parameters $b_0$ and $b_1$ are determined by
fits to the data; in the impulse approximation they are 
the isoscalar and isovector hadron-nucleon scattering lengths, 
respectively.  The factor $(1+\frac{\mu}{M}\frac{A-1}{A})$
transforms these from the two-body CM system to the $\bar p$-nucleus
CM system \cite{BFG97,GWa64}. A Coulomb potential due to the finite
size charge of the nucleus was also included in the interaction
together with vacuum polarization corrections.
The optical potential is used in a Klein-Gordon (KG) equation to calculate 
strong interaction observables to be compared with experiment. Note that
the KG equation yields to a very good approximation
the $j$-averaged results from the Dirac equation.
This is adequate as the PS209 experimental results do not distinguish
between the different $j$ values.

For the nuclear densities the $\rho _p$ may be obtained from the 
generally known charge distribution by unfolding the `finite size' of
the proton but various simplifications are required in modelling $\rho _n$.
The difference between the rms radii of the
neutron and proton distributions turned out to be
a significant factor in determining 
strong-interaction level shifts and widths in $\bar p$ atoms
\cite{FGM05,TJL01,JTL04} and these differences are parameterized by a linear
dependence on the relative neutron excess, namely
\begin{equation} \label{eq:RMF}
r_n-r_p = \gamma \frac{N-Z}{A} + \delta \; ,
\end{equation}
with $\gamma$ close to 1.0~fm and $\delta$ close to zero. Two-parameter
Fermi distributions are used for $\rho _p$ and $\rho _n$ with
the `halo' shape chosen for the latter 
where the larger $r_n$ in nuclei with neutron excess
is due to larger diffuseness parameter \cite{TJL01,FGM05}.
In addition it was shown  \cite{FGM05,Fri09}
that introducing a finite-range folding into the otherwise
zero-range potential Eq.(\ref{eq:pbarspotl}) leads
to significantly improved  agreement between calculation and experiment
in global fits to the data.

\begin{figure}[t]
 \begin{center}
  \includegraphics[height=11cm,width=0.8\linewidth]{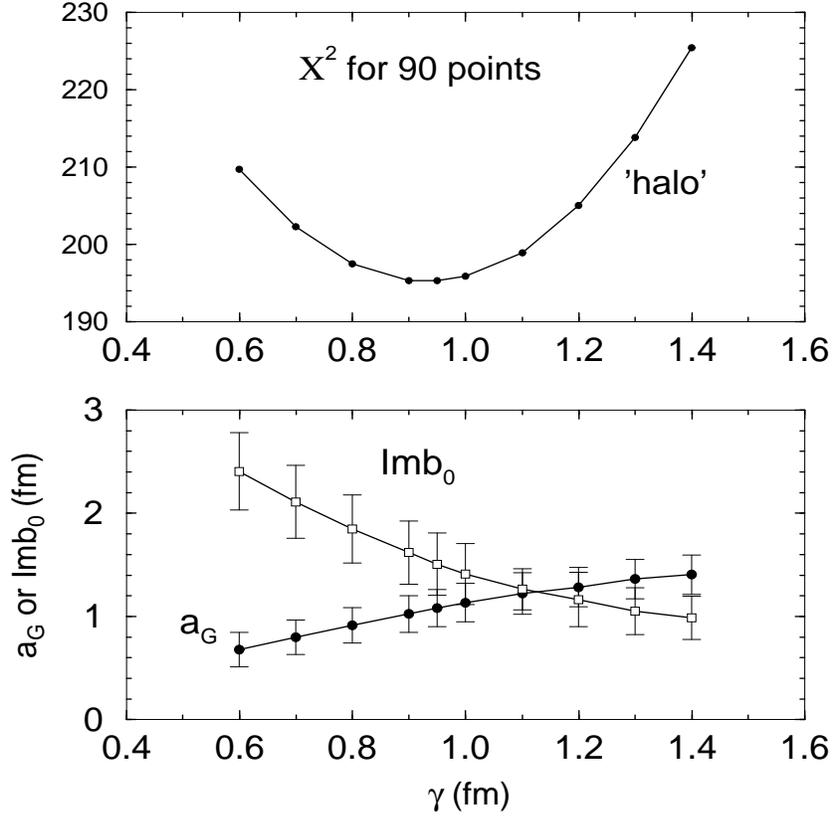}
\caption{Summary of global fits to $\bar p$ atoms as function of the
neutron density parameter $\gamma$ of Eq.(\ref{eq:RMF}). Top: $\chi ^2$
values for 90 data points, bottom: resulting parameters Im $b_0$ and $a_G$,
see text.}
  \label{fig:pbarsA}
 \end{center}
\end{figure}

Results of global fits to 90 data points for $\bar p$ atoms with a finite
range interaction are shown in fig.\ref{fig:pbarsA} using $\delta=-0.035$ fm. 
Three parameters,
Re$b_0$, Im$b_0$ and $a_G$ were adjusted for each value of the parameter
$\gamma$ of Eq.(\ref{eq:RMF}), where $a_G$ is the range parameter of a
Gaussian representing a finite range interaction with rms radius
of $ (3/2)^{1/2}a_G$. 
Almost identical 
results are obtained if a Yukawa interaction replaces the Gaussian, provided
the two have the same rms radius. Earlier work \cite{FGM05} showed that
at the minimum of $\chi ^2$ the isovector parameter $b_1$ is
consistent with zero. 
The real part of the potential plays a relatively minor role compared to
 the imaginary 
part because of the very strong annihilation of antiprotons on nuclei.
The best fit implies $\chi ^2$ per degree of freedom
$\chi ^2$/d.f.= 2.2 for 29 different nuclei between $^{16}$O
and $^{208}$Pb \cite{FGM05}.

Next we examine the $\bar p$-nucleus potential at low energies
above threshold which we wish to use for comparisons with $\bar n$-nucleus
interactions. Measurements of elastic scattering of
antiprotons by $^{12}$C, $^{40}$Ca and $^{208}$Pb were made in the
mid 80s and analyzed using standard low-energy optical model methods,
see Janouin et al. \cite{JLG86} and references therein. Here we 
show that a potential very similar to the one derived from fits 
to $\bar p$ atoms is capable of describing the elastic scattering 
of antiprotons by nuclei near 300~MeV/c
(48 MeV energy).

Above threshold the optical potential was used to calculate the complex
phase shifts $\delta_l$ for several partial waves from which the observables
were calculated. For example, the total reaction cross section which
for $\bar p$ and $\bar n$  represents, to a very good approximation,
the annihilation cross section, is given by
\begin{equation}
\sigma_R=(\pi/k^2)\Sigma (2l+1)T_l
\label{eq:sigmaR}
\end{equation}
where $k$ is the c.m. wave number and the transmission $T_l$ is given by
\begin{equation}
T_l=1-\exp (-4~{\rm Im}~\delta_l).
\label{eq:tranS}
\end{equation}

\begin{figure}[t]
 \begin{center}
  \includegraphics[height=9cm,width=0.8\linewidth]{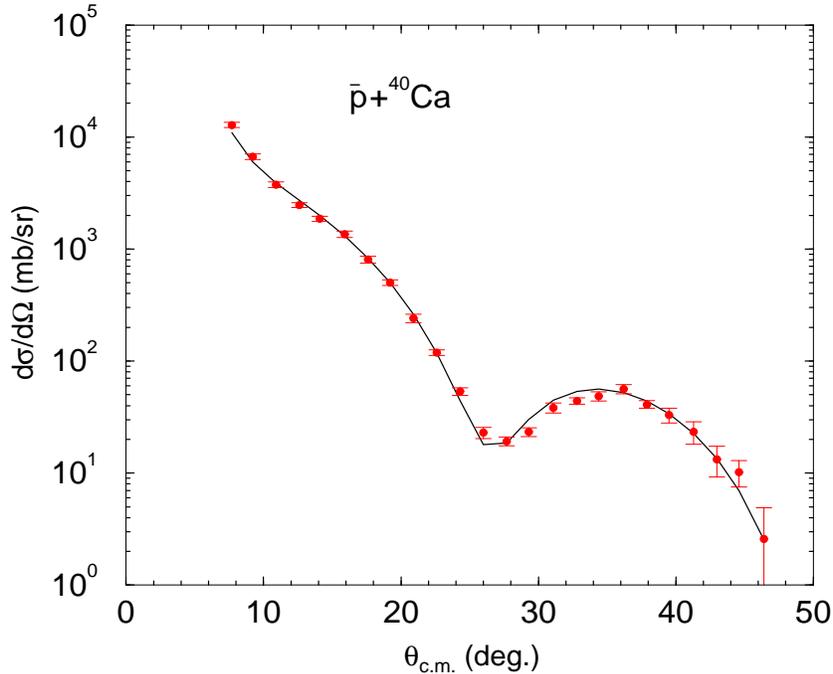}
\caption{Comparing calculation with experiment for elastic scattering
of 48 MeV antiprotons by $^{40}$Ca. Data are from \cite{JLG86}.}
  \label{fig:Ca48MeV}
 \end{center}
\end{figure}

In a first phase we considered the elastic scattering of 48 MeV 
antiprotons by $^{12}$C and $^{40}$Ca where the nuclear densities 
for $^{12}$C were of the modified harmonic oscillator type \cite{VJV87}.
For these targets  of
$N=Z$ nuclei there is no dependence on the parameter $\gamma $
 of Eq.(\ref{eq:RMF}) and the comparisons between calculation and
experiment involve only the three parameters $b_0$ and $a_G$.                 
Fits were made to the scattering data for the two targets put
together leading to $\chi ^2$ of 127 for 68 data points. The resulting
parameters are Re~$b_0=(0.40\pm0.04$) fm, Im~$b_0=(1.25\pm0.05$) fm
and $a_G=(1.34\pm0.05$) fm. Comparing with fig.\ref{fig:pbarsA} we
see agreement within errors with the Im $b_0$ and $a_G$ values at
the minimum of the $\chi ^2$ for $\bar p$ atoms. 
The same applies also to  Re $b_0$,
not shown in the figure. In a second phase we repeated the fit to
the scattering data for  $^{12}$C, $^{40}$Ca and $^{208}$Pb put
together, a total of 88 data points. For $^{208}$Pb a value of 
$\gamma$=0.9~fm was used following the best fit for $\bar p$ atoms. 
The quality of the fits and values of 
parameters remain the same as without  $^{208}$Pb, within errors.
The $\chi ^2$/d.f. of 2.2 for the three targets, relating to five
different experiments, is satisfactory. Note that only three parameters
are required to achieve this result.
Fig.\ref{fig:Ca48MeV} shows as an example the fit to the experimental
elastic scattering from $^{40}$Ca.
The overall picture hardly changes when the  Gaussian folding is replaced 
by a Yukawa folding having the same rms radius. The present analysis
can be extended up to 600~MeV/c \cite{JLG86} with very small changes
in the final results.

Further tests of the potential model above threshold can be made with
 the small number of measured total annihilation cross sections for
$\bar p$ on nuclei.
These are compared to the calculated total reaction cross sections
from the above optical potentials which could be somewhat larger than
the annihilation cross sections above the threshold for charge exchange
reactions. Table~\ref{tab:sigNe} shows calculated total
reaction cross sections with the experimental results of Bianconi et al.
\cite{BBB00} and of Balestra et al. \cite{BBB86} for Ne. Similar 
comparisons are made in table~\ref{tab:skgAD} with the recent 
results of the ASACUSA collaboration at 100~MeV/c \cite{BCH11}. It is 
seen that the overall agreement between calculations based on the above
optical model and experiment are satisfactory.

It is concluded that the interaction of antiprotons with medium-weight
to heavy nuclei from sub-threshold $\bar p$ atoms up to 600 MeV/c
is described well by an isoscalar optical potential that depends very
little on energy.

\begin{table}
\caption{$\bar p$ annihilation cross sections on Ne. Experimental 
results from refs. \cite{BBB00} and \cite{BBB86}.} 
\label{tab:sigNe}
\begin{center}
\begin{tabular}{ccccc}
\hline
p$_{lab}$(MeV/c) & 57 & 192.8 & 306.2 & 607.9 \\ \hline
$\sigma _{exp}$(mb) & 2210$\pm$1105 & 956$\pm$47 & 771$\pm$28 & 623$\pm$21 \\
$\sigma_{calc}$(mb) & 2760 & 1040 & 865 & 676 \\ \hline
\end{tabular}
\end{center}
\end{table}
\begin{table}
\caption{$\bar p$ annihilation cross sections at 100 MeV/c.Experimental
results from ref.\cite{BCH11}.} 
\label{tab:skgAD}
\begin{center}
\begin{tabular}{cccc}
\hline
target & Ni & Sn & Pt \\  \hline
$\sigma _{exp}$(mb) & 3300$\pm$1500 & 4200$\pm$900 & 8600$\pm$4100 \\
$\sigma_{calc}$(mb) & 3170 & 5560 & 8620 \\ \hline
\end{tabular}
\end{center}
\end{table}

\section{Antineutron-nucleus interaction}
\label{sec:nbarsNUC}

\begin{figure}
 \begin{center}
  \includegraphics[height=11cm,width=0.8\linewidth]{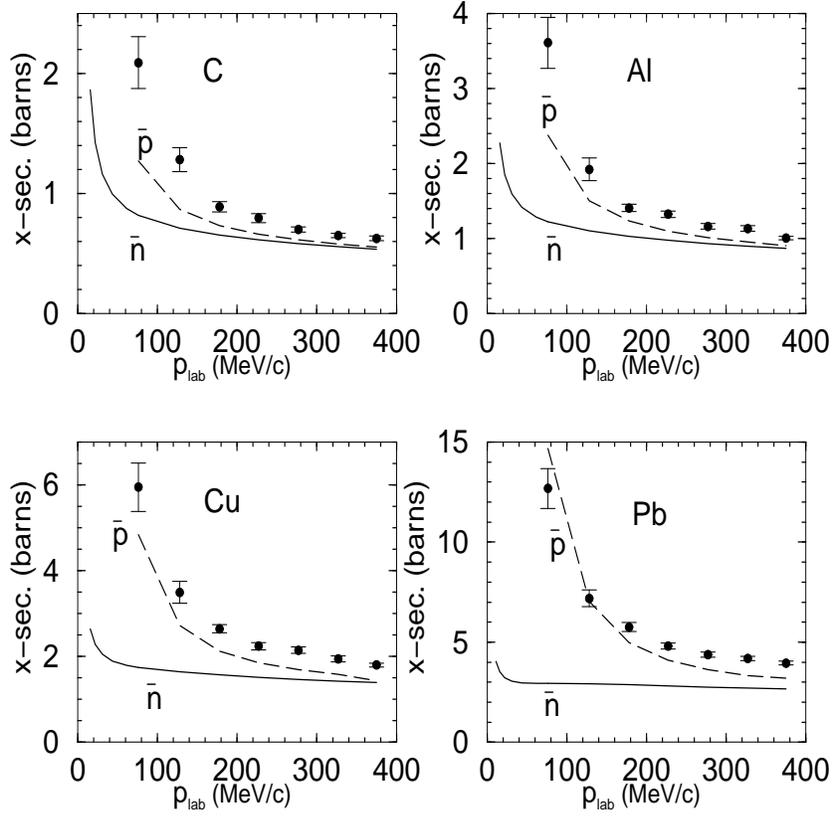}
\caption{Comparing calculation with experiment for total annihilation
cross sections  on nuclei using the potential 
derived in sec.\ref{sec:pbarsNUC}. Solid curves for $\bar n$, dashed
curves for $\bar p$. Experimental results for  $\bar n$ are 
from \cite{ABB02}.} 
  \label{fig:nbarsA}
 \end{center}
\end{figure}

Finally we turn to the annihilation cross sections of $\bar n$ on
nuclei which was the main motivation for the present work. 
For lack of comparable set of experimental results for $\bar p$-nucleus
interaction we base most of the comparisons (except for one point)
on predictions made with the optical potential. 
Having demonstrated the ability of an optical potential to produce
good fits to antiproton-nucleus observables across threshold without
a need for an isovector term, it is natural 
to  compare predictions made with the same optical potential
with the 
experimental results for antineutron-nucleus interactions. 
Fig.\ref{fig:nbarsA} shows 
(solid curves) comparison between calculations and experiment
for four out of the six targets studies by Astrua et al. \cite{ABB02}
where the calculations used the best-fit $\bar p$-nucleus potential
of sec.\ref{sec:pbarsNUC} also for antineutrons.
It is seen that whereas above 250 MeV/c there is reasonable agreement
between calculation and experiment, the calculations underestimate experiment
by up to a {\it factor} of 3-4 below 100~MeV/c. 
Moreover, calculations do not follow the trend of the experimental 
cross sections.

For lack of corresponding experimental results for antiprotons, 
fig.\ref{fig:nbarsA} shows 
(dashed curves) also calculations for $\bar p$ annihilation
cross sections on the same targets at the same energies using the optical
potential derived in sec.\ref{sec:pbarsNUC}. Compared with the 
calculated $\bar n$ cross sections the Coulomb focusing effect is
clearly seen, as expected. The start of the $1/v$ rise for $\bar n$ 
is also seen, shifting to lower and lower energies as the size of the 
nucleus increases.

 Attempts to improve the agreement between  
calculations and experiment for $\bar n$ by varying the potential
parameters failed to reduce the huge discrepancies at the lower momenta
unless the range parameter $a_G$ was increased 
from 1.3 to 3.3 fm, implying
a rms radius of 4.0 fm for the folding interaction. This value is 
significantly larger than e.g. the finite size of the proton or the 
pion Compton wavelength.

\begin{figure}[t]
 \begin{center}
  \includegraphics[height=8cm,width=0.8\linewidth]{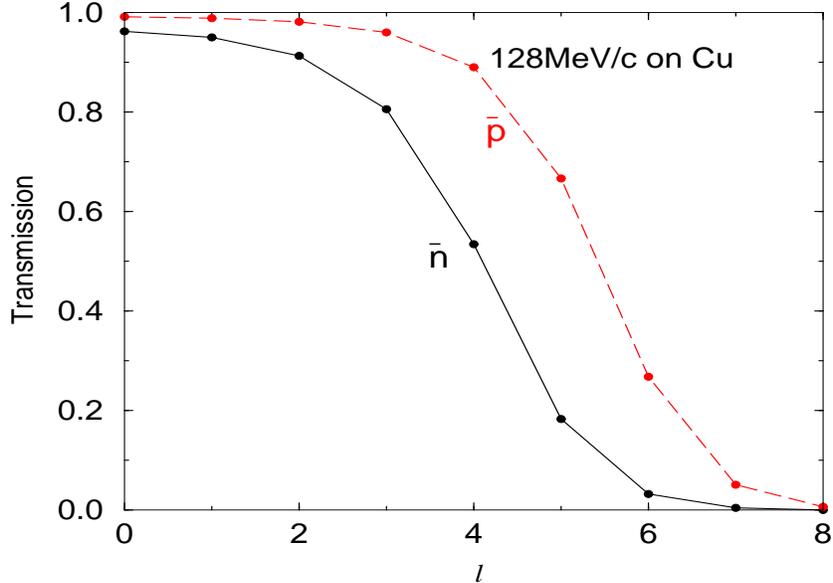}
\caption{Transmission coefficients for 128 MeV/c antineutrons
(solid curve) and antiprotons (dashed curve) on Cu 
 as function of the angular
momentum $l$ calculated from
the potential of sec.\ref{sec:pbarsNUC}.}
  \label{fig:nbarsCuT}
 \end{center}
\end{figure}

Insight into the working of the Coulomb focusing effect may be gained 
from fig.\ref{fig:nbarsCuT} showing transmission coefficients for 
128 MeV/c antineutrons
(solid curve) and antiprotons (dashed curve) on Cu
 as function of the angular
momentum $l$ calculated from
the potential of sec.\ref{sec:pbarsNUC}. 
 The larger cross sections calculated
with the Coulomb potentials included  are due to the increase in 
the number of partial waves
which contribute to the cross section with the (2$l$+1) weight,
see Eq.(\ref{eq:tranS}).

\section{Discussion}
\label{sec:summ}

Concluding that a simple global potential is capable of reproducing
well all the experimental results on strong interaction effects in
antiprotonic atoms is not new \cite{FGM05}.  We have shown here that
the same potential 
produces also agreement with mesurements
of elastic scattering of $\bar p$ by $^{12}$C, $^{40}$Ca and $^{208}$Pb
at 300~MeV/c and with the few measured annihilation cross sections
on nuclei. This isoscalar potential has been used here
to compare $\bar n$ and $\bar p$
cross sections on nuclei at very low energies.

\begin{figure}[t]
 \begin{center}
  \includegraphics[height=9cm,width=0.8\linewidth]{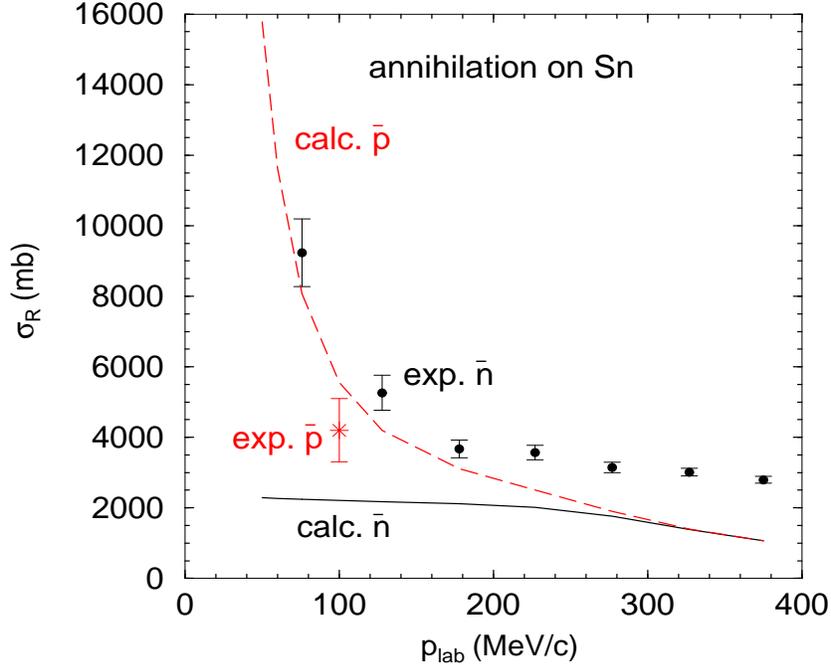}
\caption{Comparing calculation with experiment for total annihilation
cross sections  on Sn using the potential
derived in sec.\ref{sec:pbarsNUC}. Solid curve for $\bar n$, dashed
curve for $\bar p$. Experimental results for  $\bar n$ (filled circles) are
from \cite{ABB02}. The single experimental point for $\bar p$ 
(star) is from
\cite{BCH11}, see also Table \ref{tab:skgAD}.}
  \label{fig:Snall}
 \end{center}
\end{figure}

In comparing total annihilation cross sections for antineutrons on
nuclei with the corresponding values for antiprotons, one may be
guided by fig.\ref{fig:pbarnbarp} which shows that at very low
energies the $\bar p p$ cross sections exceed significantly
the $\bar n p$ ones. If the mechanism is the Coulomb focusing then
the effect is expected to become stronger as the atomic number 
of the nucleus increases. This is indeed observed with calculated
cross sections based on an optical potential which fits all the 
available experimental results for $\bar p$-nucleus interactions
across threshold. Unfortunately comparable experimental results 
for $\bar p$ and $\bar n $ exist
only for the proton as a target and we had to use
optical potentials for comparisons with the $\bar n$-nucleus
experimental results. However, there is a single experimental result
which could be useful in this respect, namely, a preliminary result
of the ASACUSA collaboration for the total annihilation
cross section of $\bar p$ on Sn at 100 MeV/c \cite{BCH11}. Table 
\ref{tab:skgAD} shows it to agree with the predictions of the optical
potential and in fig.\ref{fig:Snall} we see it in relation to the 
$\bar n$ cross sections on the same target. The figure suggests that
the $\bar n$  cross sections are larger than the corresponding $\bar p$
cross sections, contrary to the evidence from 
fig. \ref{fig:pbarnbarp}.

 A possible conclusion that the $\bar n$-nucleus total
annihilation cross sections at very low energies are equal to or 
larger than the corresponding $\bar p$ cross sections will be at variance
with expectations based on smooth optical potentials and on experimental
results for the proton as a target. A theoretical approach that may
explain all the observations has not been presented so far. However, an
experimental approach to this `puzzle' is possible in the foreseeable
future by measuring total annihilation cross sections
for antiprotons on the six nuclear targets of Astrua et al. \cite{ABB02}
at the same energies.

\section*{Acknowledgements}
I wish to thanks T. Bressani for useful discussions and for providing
in numerical form the detailed results of ref. \cite{BBC97}. 
Discussions with A.~Gal and A. Leviatan are gratefully acknowledged.


\vspace{30mm}


\begin{thebibliography}{99}

\bibitem{ABB02} M. Astrua et al., Nucl. Phys. A 697 (2002) 209.



\bibitem{FGa07} E. Friedman, A. Gal, Phys. Rep. 452 (2007) 89 and references
                therein.

\bibitem{GPT68} G.W. Greenlees, G.J. Pyle, Y.C. Tang, Phys. Rev. 171 (1968) 1115. 

\bibitem{BBC96} A. Bertin et al., Phys. Lett. B 369 (1996) 77.

\bibitem{ZBB99} A. Zenoni et al., Phys. Lett. B 461 (1999) 405.

\bibitem{Ben97} A. Benedettini et al., Nucl. Phys. B (Proc. Suppl.) 56A
(1997) 58.

\bibitem{BBC97} A. Bertin et al., Phys. Lett. B 410 (1997) 344. 

\bibitem{BFG01} C.J. Batty, E. Friedman, A. Gal, Nucl.Phys. A 689 (2001) 721.



\bibitem{TJC01}A.~Trzci\'{n}ska et al., Nucl. Phys. A  692 (2001) 176c.

\bibitem{FGM05} E. Friedman, A. Gal, J. Mare\v{s}, Nucl. Phys. A  761.
(2005) 283.

\bibitem{BFG97} C.J. Batty, E. Friedman, A. Gal, Phys. Rep. 287 (1997) 385.

\bibitem{GWa64} M.L. Goldberger, K.M. Watson, Collision Theory, John
Wiley, New York, 1964.

\bibitem{TJL01}A.~Trzci\'{n}ska et al., Phys. Rev. Lett. 87 (2001) 082501.

\bibitem{JTL04} J. Jastrz\c{e}bski et al.,  Int. J. Mod. Phys. E
 13 (2004) 343.

\bibitem{Fri09} E.~Friedman, Hyperfine Interact  193 (2009) 33.

\bibitem{JLG86} S.~Janouin et al., Nucl. Phys. A 451 (1986) 541.

\bibitem{VJV87} H. de Vries et al., At Data Nucl. Data Tables 36 (1987) 495.

\bibitem{BBB00} A. Bianconi et al., Phys. Lett. B 481 (2000) 194.

\bibitem{BBB86} F. Balestra et al., Nucl. Phys. A 452 (1986) 573.

\bibitem{BCH11} A. Bianconi et al., Phys. Lett. B 704 (2011) 461.









\end{thebibliography}
\end{document}